\documentclass[
nopreprint,
]{jasatex}

\usepackage{graphicx}
\usepackage{dcolumn}
\usepackage{amsmath,amsfonts}
\usepackage{bm}

\begin{document}


\title[Stability of noise-based correlation monitoring]
{Stability of Monitoring Weak Changes in  Multiply Scattering Media with Ambient Noise Correlation: Laboratory Experiments.}

\author{C\'eline Hadziioannou, Eric Larose, Olivier Coutant, Philippe Roux and Michel Campillo}
\affiliation{Laboratoire de G\'eophysique Interne et Tectonophysique, CNRS \& Universit\'e J. Fourier, BP53, 38041 Grenoble, France. Email: eric.larose@ujf-grenoble.fr}

\date{\today}

\begin{abstract}
Previous studies have shown that small changes can be monitored in a scattering medium by observing phase shifts in the coda. Passive monitoring of weak changes through ambient noise correlation has already been applied to seismology, acoustics and engineering. Usually, this is done under the assumption that a properly reconstructed Green function as well as stable background noise sources are necessary. In order to further develop this monitoring technique,  a laboratory experiment was performed in the 2.5MHz range in a gel with scattering inclusions, comparing an active (pulse-echo) form of monitoring to a passive (correlation) one. Present results show that temperature changes in the medium can be observed even if the Green function (GF) of the medium is not reconstructed. Moreover, this article establishes that the GF reconstruction in the correlations is not a necessary condition: the only condition to monitoring with correlation (passive experiment) is the relative stability of the background noise structure.
\end{abstract}

\maketitle

\section{Introduction}
In order to image a complex medium the impulse response, or Green function (GF), of that medium is needed. Classically, the impulse response is retrieved by active means: a signal generated at one point (\textit{eg.} an earthquake) is recorded at another (a passive receiver like a seismic station), and this record is treated as the band-pass filtered GF. Over the last fifteen years, developments in helioseismology\cite{duvall1993} and in acoustics\cite{weaver2001,lobkis2001} showed that information about a medium can be extracted from diffuse (coda) waves or ambient background noise. Since then, seismologists are turning to this \textit{passive imaging} technique\cite{larose2006c}. In this latter case the seismic coda\cite{campillo2003} or seismic noise\cite{shapiro2004} is correlated to reconstruct the GF, either by averaging over space, time, or both. \textit{Passive imaging} requires some assumptions: one needs uniformly distributed noise sources and/or long enough record duration to get the correlation function to converge to the GF. \\

Monitoring dynamic media is a separate issue and, as we will see, is based on different (and weaker) assumptions.
In the 80's, Poupinet et al.\cite{poupinet1984} proposed to use coda waves to monitor velocity changes in scattering media (the after-mentioned \textit{doublet} technique). At first glance, coda waves might appear as a jumbled mess of wave arrivals. In fact, they consist of the waves which have described long, scattered paths through the medium, thereby sampling it thoroughly. As a result, these scattered waves are more sensitive to small variations than the ballistic waves. 
For this reason, the information in the coda can be exploited to monitor small changes in a medium. 
This technique, analogous to Diffuse Wave Spectroscopy (DWS\cite{pine1988}) in optics, was later named \textit{coda wave interferometry} (CWI\cite{snieder2002}). It tracks the tiny phase changes in the coda that are caused by velocity changes in the medium. A major issue of the \textit{doublet} technique is that it requires stable reproducible sources, which are hardly available in seismology. Thus, a more recent idea was to combine noise-based \textit{passive imaging} with the \textit{doublet} technique\cite{sabra2006,sens-schoenfelder2006,sens-schoenfelder2008,brenguier2008a}. First, one correlates the background noise between two receivers. Second, one analyzes small phase changes at large lapse times (coda) in the correlations. This forms the basis of \textit{passive monitoring} (or passive image interferometry\cite{sens-schoenfelder2006}) with seismic noise.   Noise based \textit{passive monitoring} seems to simultaneously require two conditions. First: a homogeneous distribution of sources in space, and second: temporal stability of these sources.\\

In seismology, most of the noise (between 0.01 and 1 Hz) generated in the oceans\cite{rhie2004,stehly2006}   shows strong spatio-temporal variabilities. This feature is in favor of \textit{passive imaging} as long as records are long enough (duration of the order of a year) to average over a large distribution of sources. But to \textit{passively monitor} dynamic phenomena over a few days or less, this feature seems very unfavorable. In this paper, we investigate the effect of these non-ideal conditions on the reliability of \textit{passive monitoring}. We will also examine if the GF reconstruction in the correlation is a necessary condition to perform \textit{passive monitoring}. To that end, we test the \textit{passive monitoring} technique under degraded conditions in a controllable (laboratory) environment. In section II of the present paper, we describe the experimental setup and our motivations to do small-scale seismology. 
In section III, we compare two different data processing procedures to extract velocity variations from the coda wave. One is the \textit{doublet} technique introduced twenty years ago. The other, referred to as \textit{stretching}, was developed more recently\cite{lobkis2003,sens-schoenfelder2006,larose2009}. Advantages and drawbacks of both procedures are discussed. We also investigate the robustness of these procedures when noise is introduced in the signal.
In section IV, we investigate if \textit{passive monitoring} is still possible when the GF is not properly reconstructed     in the correlations.
Finally, in section V, we test the robustness of \textit{passive monitoring} in the case of temporally changing distribution of sources.

\section{Methodology}

\subsection{Motivations for doing analogous laboratory experiments}

\begin{table}[htbp]
	\centering
\begin{tabular}{c | c c }
			  &Seismology&Ultrasound     \\ \hline \hspace{.4cm}
Wavelength& km& mm\\
Frequency& mHz - Hz& MHz           \\
Total size&$10^{3}$ km&m\\
Total duration &month-year&min\\

\end{tabular}
\caption{Comparison of the physical parameters between seismology and ultrasound.}\label{table}
\end{table}

Seismology is based on the observation and processing of natural vibrations. In a passive field experiment where seismic waves originate from earthquakes, scientists are facing two simultaneous problems. They neither know the source location with sufficient precision, the source mechanism nor the medium of propagation. It is therefore very complex to image the source and the medium at the same time. By reproducing some features of the seismic propagation in the lab and employing controlled sources and sensors, we can focus our efforts on the physics of the wave propagation and develop new methods more comfortably. In laboratory-scale seismology, we control for instance the size of the medium, the scattering properties, and the absorption. We are then able to adjust one parameter at a time and test the physical models and imaging techniques we develop. But the main reason for carrying out analogous ultrasonic experiments is more tactical: it is related to the order of magnitude of the physical parameters as recalled in Table~\ref{table}. Ultrasonic wavelengths are on the order of a millimeter, meaning that experiments are physically easy to handle. Additionally, the duration of a single ultrasonic experiment is very short (one minute) compared to seismology where we have to wait for earthquakes (year). This characteristic allows us to achieve many more experiments in the lab, and test many parameters over a wide range of magnitudes. In the view of testing processing technique to monitor weak changes, it is also of first importance to perfectly control the origin of the change in the medium. This is quite convenient in the laboratory, but almost impossible in a natural environment. These are the reasons why several seismology laboratories have decided to develop analogous experiments for methodological developments \cite{snieder2002,fukushima2003,vanwijk2004,larose2005b}. Our article presents one analogous ultrasonic experiment not only devoted to the study of the physics of wave propagation in heterogeneous media, but also to the development of new techniques applicable to seismic waves in geosciences.

\subsection{Scattering properties of the medium}
\begin{figure}[htbp]
	\centering
		\includegraphics[width=8cm]{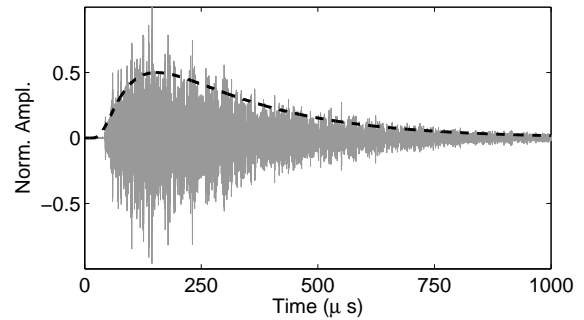}
		\caption{Gray line : acoustic field transmitted through the 64~mm thick bubble-gel mixture (in normalized amplitude). Black broken line: 2D diffusion equation for $\ell^{\star}=3.5$~mm and $\ell_a=200$~mm. }
	\label{fig1}
\end{figure}

We perform the experiment on a 80~mm$\times$64~mm block of Agar-Agar gel which consists of 95\% water and 5\% Agar (by weight). 8.5\% of the volume of the gel consists of small air bubbles, with diameters between 100 $\mu$m and 1 mm. These bubbles render the medium multiply scattering. The source emits a pulse at 2.5~ MHz (100\% frequency bandwidth). For simplicity, we neglect the electronic noise in the experiment. Since shear waves are strongly attenuated, we assume that only P-waves are propagating in the medium and are eventually recorded.
To estimate the scattering properties of the medium, we performed several experiments in the transmission configuration for several medium thickness.     
From the attenuation of coherent plane wave, we obtain an estimation of the elastic scattering mean free path averaged in the working frequency band: $\ell_e\approx 3.5 mm$. Since the scatterers' size is smaller or of the order of the wavelength, scattering is isotropic and we expect a transport mean free path $\ell^{\star}$ of the same order. In Fig.~\ref{fig1} we plot an example of a diffuse record transmitted through 64~mm of our heterogeneous medium. A theoretical fit is obtained from the two-dimensional (2D) diffusion equation (including reflections from the sides) and plotted as a broken black line. The diffusion constant is $D=v_P \ell^{\star}/2$ and we assume     $\ell_e\approx\ell^{\star}$. The absorption length $\ell_a$ is the fit parameter, the best fit is obtained for $\ell_a=200$~mm. This corresponds to an absorption three times stronger than in pure water \cite{zagzebski1996} which is due to dissipation by the agar material.

\section{Comparison of data processing techniques}

Two processing techniques have been proposed in the literature to estimate relative velocity changes $dV/V$ in the diffuse coda. The first one, called the seismic \textit{doublet} technique, was developed for geophysical purposes about twenty years ago \cite{poupinet1984}. The idea is to measure a time-shift between two different records in limited time-windows centered at $t$ in the coda. By repeating this procedure at different times $t$, it is possible to plot the delay $\delta t$     versus $t$. The velocity variation is simply the average slope of $\delta t (t)$: $dV/V=-\delta t/t$. Doing so, we implicitly assume that the time-shift is constant within the considered time-window, which might be not the case. This processing found remarkable applications in geophysics, including recent developments in volcano eruption prediction \cite{brenguier2008a} and active fault monitoring \cite{brenguier2008b}.\\

Another idea\cite{lobkis2003,sens-schoenfelder2006} is to interpolate the coda at times $t(1- \varepsilon)$ with various relative velocity changes $ \varepsilon$. This corresponds to stretching the time axis. The actual velocity change is obtained when the interpolated coda best fits the original data.
Because we do not assume a constant time-shift in the considered time window [0 T], we can process the whole data at once,     which is expected to result in a more stable, and thus more precise, estimation of $dV/V$. One drawback is that this latter processing assumes a linear behavior $\delta t (t)$ versus $t$, or a constant relative velocity change $dV/V= \varepsilon$, which is sometimes not the case in complex heterogeneous media. No quantitative comparison between these two techniques have been established in the literature. In the following section, we propose to test both techniques on the same data set, and analyze their sensitivity to the SNR of the records.



\subsection{Active experiment: high quality data.}

\begin{figure}
	\centering
                    \includegraphics[ width=8cm]{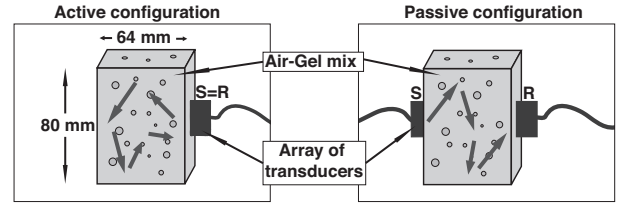}
                    \caption{Experimental setup. Left: the active experiment in the pulse-echo configuration. Right: the passive experiment in the transmission configuration before auto-correlation.} 
                    \label{FigSetup}
\end{figure}

\begin{figure}
	\centering
                    \includegraphics[ width=8cm]{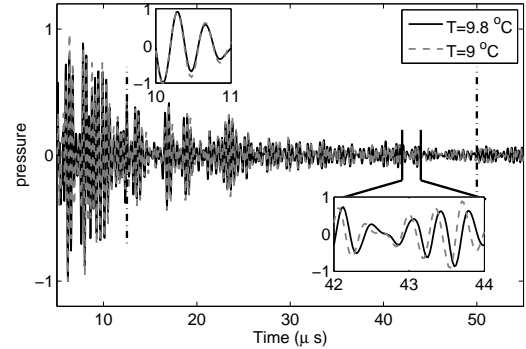}
                    \caption{Example of two records $h_0(t)$ and $h_3(t)$ acquired at the same position (same transducer) but at two different dates. Between the two acquisitions, the temperature has increased by $0.8^{\circ}$C, which is hardly visible in the early part of the record (inset between 10 and 11~$\mu$s) but very clear in the late coda (inset between 42 and 44~$\mu$s).} 
                    \label{fig3}
\end{figure}

In this experiment, we attach a set of transducers on one side of the gel which act as both sources and receivers of the signal (figure \ref{FigSetup}-left). The source emits a 2.5~MHz pulse. The signal is collected on the same transducer (R) in the pulse-echo configuration.     This procedure is repeated on 7 different channels. As the gel contains a large amount of scatterers, the emitted waveform is multiply scattered before reaching the transducer again. 
A typical ultrasonic record is plotted in Fig.~\ref{fig1}. Note that the early 5$\mu s$ are muted for technical reasons. This record is composed of the GF   of the air-gel mix sample $G(R,R,t)$ and the source wavelet $e(t)$:
\begin{equation}\label{h0}
h_0(t)=G_0(R,R,t)\otimes e(t)
\end{equation}
where $\otimes$ stands for convolution. This experiment is repeated 4 times while the temperature of the medium slowly increases by     about 0.8$^{\circ}$C, as measured by a digital thermometer placed beneath the gel.     
We assume that the first effect of a temperature change is to stretch the record in time by $ \varepsilon_k=dV/V$, and to additionally slightly distort it\cite{lobkis2003}.     This weak distortion, noted $f(t)$, is not studied here, although it contains precious information about the medium and its evolution. An example of two records is displayed in Fig.~\ref{fig3}. After a (small) temperature change, the record rewrites:

\begin{eqnarray}\label{hk}
h_k(t)&=&G_k(R,R,t)\otimes e(t)\\
&=&\left[G_0\left(R,R,t\left(1+ \varepsilon_k\right)\right)+f(t)\right]\otimes e(t)
\end{eqnarray}


For each temperature $k$, the record $h_k(R,R,t)$ is compared to the reference waveform $h_0(R,R,t)$ to evaluate the relative velocity change in the gel sample.     Two processing techniques have been proposed in the literature to estimate $dV/V$: the \textit{doublet} technique and the \textit{stretching} technique.

\subsubsection{Doublet technique}

\begin{figure}
	\centering
                    \includegraphics[ width=8cm]{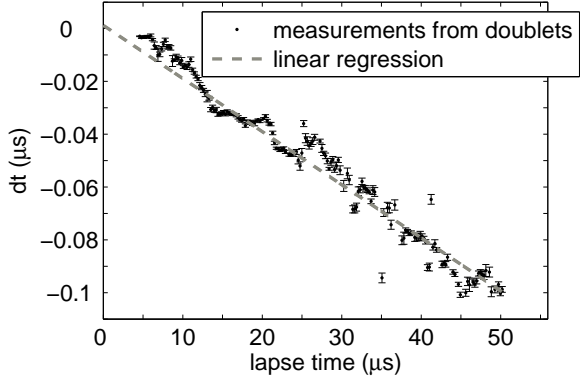}
                    \caption{Delay time evaluated for different lapse times in the coda from the \textsl{doublet} code. The broken line is the linear trend whose slope yields $dV/V$ between $k=0$ and $k=3$.} 
                    \label{doublets}
\end{figure}

The doublet technique also known as cross-spectral moving-window technique (CSMWT\cite{frechet1989}), computes the phase shift between records for consecutive, overlapping time windows. For a given window, the time shift is assumed to be constant and is estimated in the frequency domain by measuring the Fourier cross-spectrum phase. This estimator uses an accurate, unbiased Wiener filter technique \cite{jenkins1969} and produces an estimate whose confidence interval is controlled by the coherence values in the frequency band used for the analysis. The method can then measure arbitrary time-shifts between two records with enough similarity (or coherence). The key parameter in this analysis is the Fourier transform window length. The length choice is a trade-off between shift estimate accuracy, and the time resolution of possible temporal variations.\\

We use this \textsl{doublet} technique to compute the time shift between the records. The time shift between the two different records is measured in the coda between 12.5 and 50 $\mu s$. If the velocity changes homogeneously in the medium, the propagation time will vary proportionally to the propagation distance, producing a phase shift between records varying linearly with lapse time. The relative velocity change can be retrieved by measuring the slope of the phase shift as a function of lapse time, as shown in figure \ref{doublets}.

\subsubsection{Stretching interpolation technique}

\begin{figure}
	\centering
                    \includegraphics[ width=8cm]{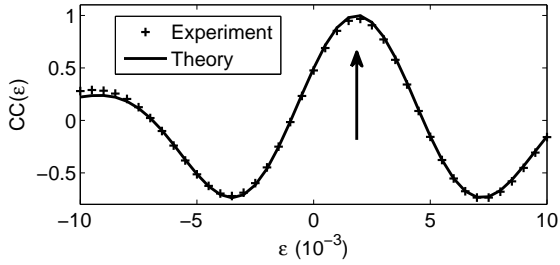}
                    \caption{The correlation coefficient $CC( \varepsilon)$ is evaluated at two temperatures for $k=$0 and 3 in the $\left[12.5~\mu s - 50~\mu s\right]$ range. The maximum, obtained for a relative velocity change of $ \varepsilon_3=1.86~ 10^{-3}$, is indicated by the vertical arrow. It corresponds to an increase of temperature of $0.8^{\circ}$C. Theory is from Eq.~\ref{eqCCnu}.} 
                    \label{CCnu}
\end{figure}

In the \textit{stretching} technique, the coda $h_k(R,R,t)$ is interpolated at times $t(1- \varepsilon)$ with various relative velocity changes, $ \varepsilon$, in the $\left[t_1 - t_2\right]$ time window. $ \varepsilon_k$ is therefore the $ \varepsilon$ that maximizes the cross-correlation coefficient:
\begin{equation}\label{CCk}
CC_k( \varepsilon)=\frac{\int_{t_1}^{t_2}h_k\left[t(1- \varepsilon)\right]h_0[t]dt}{\sqrt{\int_{t_1}^{t_2}h_k^2\left[t(1- \varepsilon)\right]dt. \int_{t_1}^{t_2}h_0^2[t]dt}}
\end{equation}

An example of correlation coefficient is plotted in Fig.~\ref{CCnu}. If we assume that $h_0$ and $h_k$ are stationary waveforms \footnote{The stationary assumption is done for the sake of simplicity of the calculation. Nevertheless, the main conclusions of the article also apply to non-stationary waveforms as decaying coda.}  and are well described by Eqs.~\ref{h0}\&\ref{hk}, we have a theoretical estimation of $CC$:
\begin{widetext}
\begin{equation}\label{eqCCnu}
CC_k( \varepsilon)=\frac{A}{\int_{\Delta \omega} \rho(\omega) d\omega} \int_{\Delta \omega}\frac{ \rho(\omega) sin\left( \omega   \varepsilon t_2\right) -sin\left( \omega   \varepsilon t_1\right)d\omega}{\omega \varepsilon \left(t_2-t_1\right)} +B( \varepsilon)
\end{equation}
\end{widetext}
which in the simple case of $t_1=0$ and $t_2=T$ simply reduces to:

\begin{equation}
CC_k( \varepsilon)=A \frac{\int_{\Delta \omega} \rho(\omega) sinc     \left(\omega \left( \varepsilon- \varepsilon_k\right) T\right)        d\omega}{\int_{\Delta \omega} \rho(\omega) d\omega} +B( \varepsilon)
\end{equation}
with $\omega$ the pulsation, $\Delta \omega$ the bandwidth, $\rho(\omega)$ the power spectrum density. The constant $A$ depends on the variance of $G$, noted $\left\langle G^2\right\rangle$ and the variance of the additional fluctuations, noted $\left\langle f^2\right\rangle$,
\begin{equation}\label{EqA}
A=\frac{\sqrt{\left\langle G^2\right\rangle}}{\sqrt{\left\langle G^2\right\rangle+\left\langle f^2\right\rangle}}
\end{equation}

$B( \varepsilon)$ is a random process of zero mean     and standard deviation\footnote{see EPAPS document \# for a detailed calculation. For more inforamtion on EPAPS, see \textit{http://www.aip.org/pubservs/epaps.html}}:
\begin{equation}\label{EqB}
\sqrt{\left\langle B^2\right\rangle}=\sqrt{\frac{2\pi}{T\Delta \omega}}\frac{ \sqrt{\left\langle f^2\right\rangle}}{\sqrt{\left\langle G^2\right\rangle+\left\langle f^2\right\rangle}     }
\end{equation}
The term containing the $sinc$ function is represented by the crosses in Fig.~\ref{CCnu}\&\ref{CCnuSNR}, and the fluctuations ($\sqrt{\left\langle B^2\right\rangle}$) around this average are in gray.
If the amplitude of the $sinc$ function is much greater than the fluctuations, $A \gg \sqrt{\left\langle B^2\right\rangle}$, the maximum of the cross-correlation coefficient $CC_k$ is obtained for $ \varepsilon= \varepsilon_k$, which provides the relative velocity change for the given state $k$. It is interesting to note that the peak of the $sinc$ function is visible even if the distortion $f(t)$ (or electronic noise noise $n(t)$, see next subsection) are strong. In such a case, increasing the integration time $T$ or the frequency bandwidth $\Delta \omega$ can reduce the fluctuations $B$. This is a crucial advantage of the present technique compared to the \textsl{doublet} technique in the case of noisy or distorted data.   \\

\subsection{Active experiment: low quality data}

\begin{figure}
	\centering
                    \includegraphics[ width=8cm]{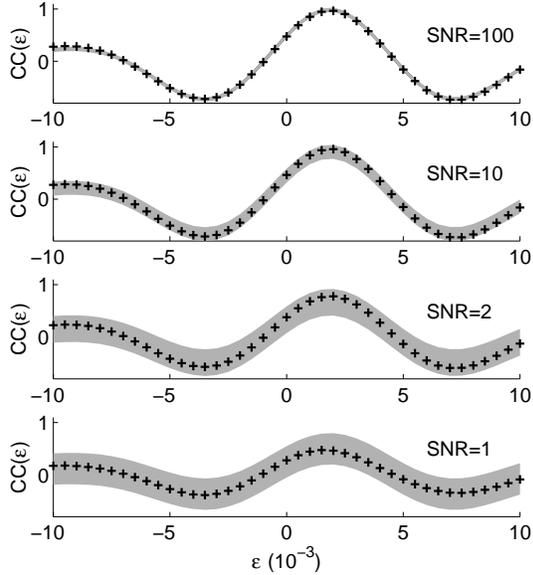}
                    \caption{The correlation coefficient $CC( \varepsilon)$ is evaluated at two temperatures $k=0\&3$ in the $\left[12.5~\mu s -50~\mu s\right]$ range, for various signal-to-noise ratio SNR. Crosses indicate experimental data. The gray background indicates expected fluctuations (Eq.~\ref{eqBSNR}) around the theory (Eq.~\ref{eqCCnuSNR}). The proper velocity change $ \varepsilon_3$ is found in all cases, though a slight difference is visible for SNR=1.} 
                    \label{CCnuSNR}
\end{figure}

To mimic a practical situation where data includes additional noise (instrumental or electronic), we add a random $\delta$-correlated noise $n(t)$ of zero mean to the signals $h(t)$ in Eq.~\ref{h0}\&\ref{hk}. For simplicity, we neglect the distortion $f_k$ in the considered time-window, and assume a stationary noise:
\begin{equation}
\left\langle n_0^2\right\rangle=\left\langle n_k^2\right\rangle=\left\langle n^2\right\rangle
\end{equation}
then we get a similar expressions as Eq.~\ref{EqA}\&\ref{EqB}, with:
\begin{equation}\label{eqCCnuSNR}
A=\frac{\left\langle h^2\right\rangle}{\left\langle h^2\right\rangle+\left\langle n^2\right\rangle}
\end{equation}

and

\begin{equation}\label{eqBSNR}
\sqrt{\left\langle B^2\right\rangle}=\sqrt{\frac{2\pi}{T\Delta \omega}}\frac{\sqrt{\left\langle n^2\right\rangle+2\left\langle n^2\right\rangle\left\langle h^2\right\rangle}}{\left\langle h^2\right\rangle+\left\langle n^2\right\rangle}
\end{equation}

\begin{widetext}
\begin{figure*}
        \begin{center} 
                    \includegraphics[ width=7cm]{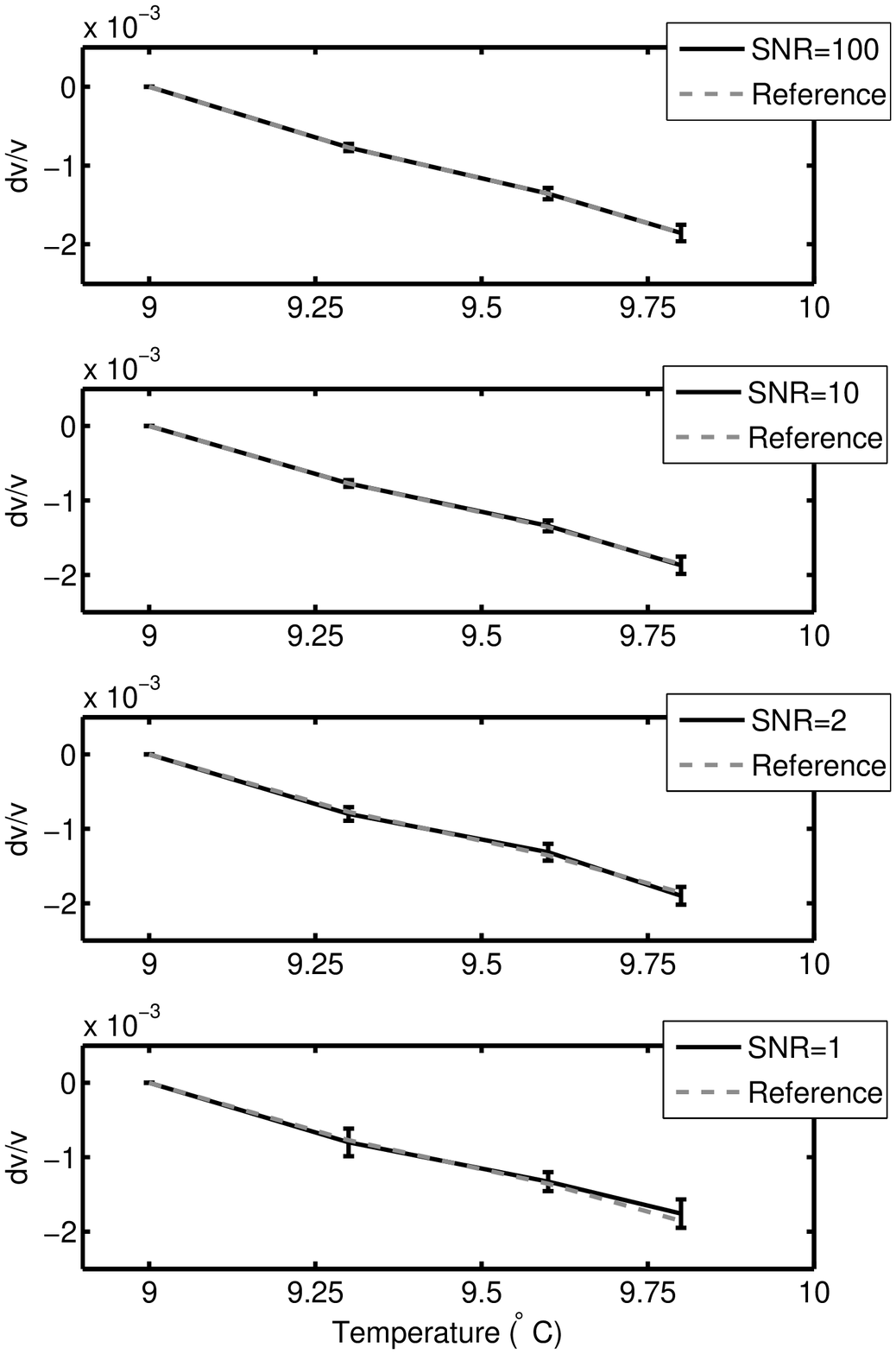}
\includegraphics[ width=7cm]{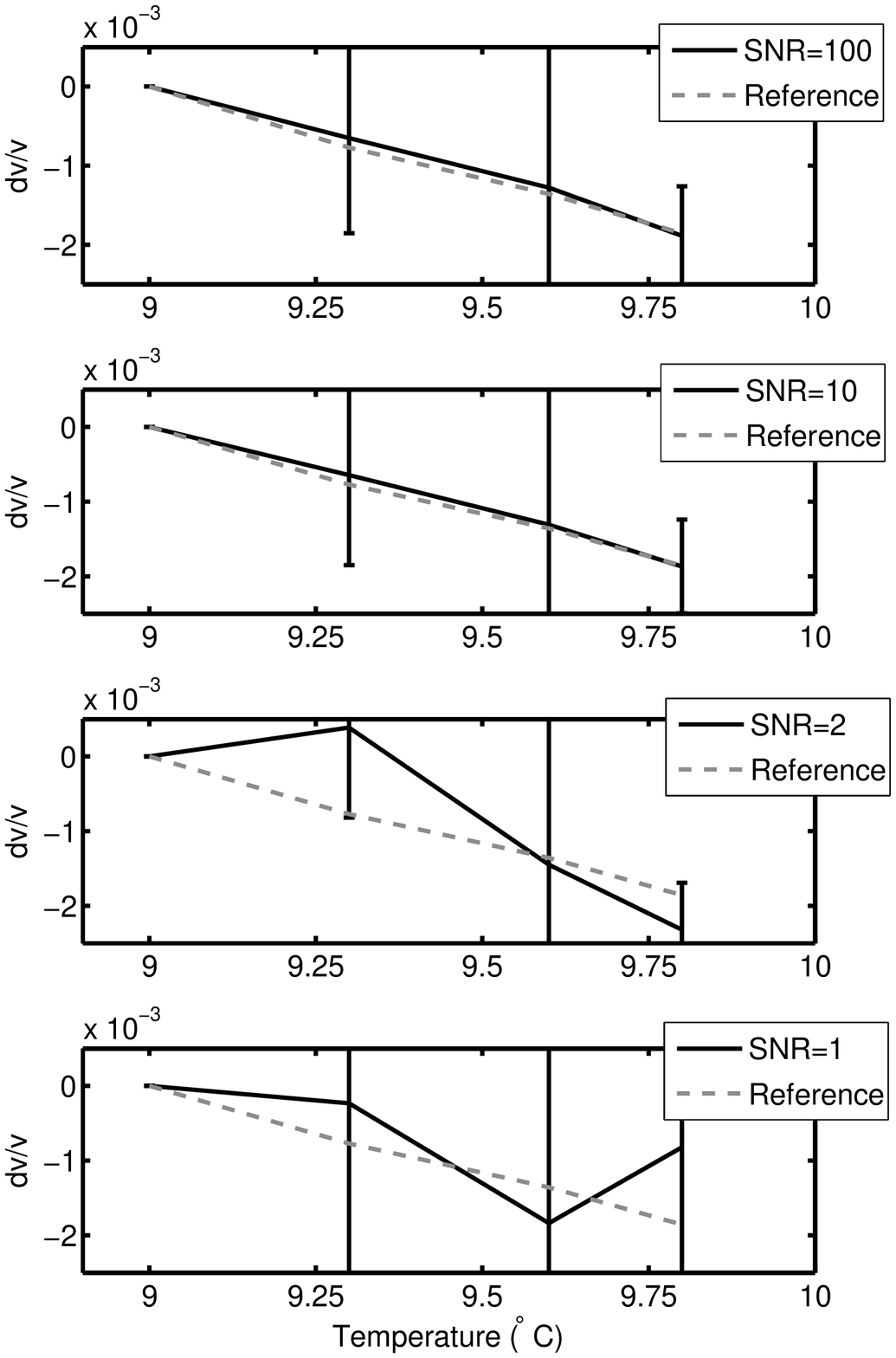}
  \newline \textit{Stretching} technique $\qquad \qquad \qquad \qquad$ \textit{Doublet} technique
                    \caption{Relative velocity changes evaluated from the \textit{stretching} (left) and for the \textit{doublet} (right) technique for various SNR. The actual $dV/V$ is satisfiyingly retrieved for any SNR$\geq 1$ with the first technique, not with the second.}
                    \label{SNRdoublets}
        \end{center}
\end{figure*}
\end{widetext}

The velocity change is measured again for signal-to-noise (SNR) ranging from 1 to 100 (Figs.~\ref{SNRdoublets}).
For a SNR of 100 and 10, we find the same results for the \textit{stretching} and for the \textit{doublet} technique. However, if the SNR is decreased to 2, the velocity variations measured from the \textit{doublet} technique are not accurate at all, while they remain relevant with the stretching technique. This establishes the \textsl{stretching} technique as a more stable processing procedure for noisy records.   Note that the connection between the fluctuation of the waveforms and the error in the estimation of $dV/V=\varepsilon$ will be subject to a separate communication. Nevertheless, this error can be visually estimated by the gray area around the theoretical curve (crosses) in Fig.~\ref{CCnuSNR}.

\subsection{Advantages and drawbacks of both techniques}
The \textit{doublet} technique (CSMWT) has been used successfully for more than 20 years to efficiently retrieve small velocity changes in the medium\cite{poupinet1984,snieder2002,brenguier2008a}. This technique does not suffer from change in amplitude of the waveform, including the coda decay, and the processing is very fast. It also manages clock errors in origin time without further processing, which is a central issue in active and passive field experiments\cite{stehly2007b}. It also allows to select a given time window in the data.\\

The \textit{stretching} technique is more recent. It is based on a grid-search for $\epsilon$, and is found to be slightly more time consuming in terms of computer processing. A noticeable disadvantage of this latter technique is also that it assumes a linear stretching of the waveform, which is not valid for media with heterogeneous changes (including the Earth). The main interest of the \textit{stretching} technique versus the \textit{doublet} one is its stability toward fluctuations (noise) in the data, as mentioned in the previous section and demonstrated by fig.~\ref{SNRdoublets}. This provides an opportunity to increase the sensitivity of detection of weak changes in the earth's crust with seismic waves\cite{brenguier2008a,brenguier2008b}.

\section{Active and passive experiment: monitoring with the correlation?}
Most previous authors suggested that monitoring weak changes in the earth with ambient seismic noise correlation is based on the assumption that those correlations yield the actual GF of the medium. Thus, the late part of the correlation is interpreted as the coda of the reconstructed impulse response. Is this assumption actually necessary to monitor the changing earth with good accuracy? We address the question in the present section by comparing relative velocity changes measured either in the autocorrelations of records from distant noise sources (the passive experimental setup), or in pulse-echo data (the active experimental setup).\\

\begin{figure}[!htbp]
        \begin{center}
                    \includegraphics[ width=8cm]{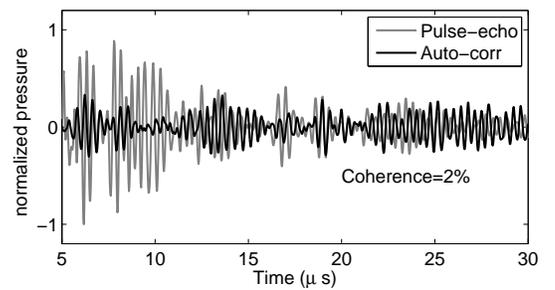}
                    \caption{Comparison of the pulse-echo data $h(R,R,t)$ obtained in the active experiment and the average autocorrelation $\partial_t h(S,R,t)\times h(S,R,t)$.} 
                    \label{Signals}
        \end{center}
\end{figure}

In the passive experiment, transducers are attached at opposite sides of the gel (figure \ref{FigSetup}-right). On one side, 16 sources (S) are acting to mimic a distribution of noise sources. Impulse responses are recorded by the 7 receiving transducers (R) and consecutively convolved by a white noise to mimic acoustic (or seismic) ambient noise. Then they are autocorrelated at each receiver. Note that the precise knowledge of the noise source position is unnecessary: the source position has no effect on the velocity change estimation. If the noise sources were uniformly distributed and the coda records long enough, these correlations averaged over sources should result in the GF for the medium \cite{weaver2001}. However, the records used in this experiment are of finite duration, and the auto-correlation has not converged to the Green function yet. This can be seen visually in figure \ref{Signals}, where the autocorrelation is plotted alongside the (time derivative of the) reference   Green function obtained in the active experiment. The fact that the two signals are uncorrelated is confirmed by the low value of the coherence between them (2\%).


\begin{figure}[!hbp]
        \begin{center}
                    \includegraphics[ width=8cm]{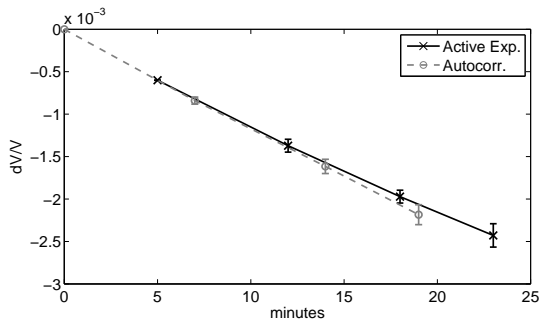}
                    \caption{Relative velocity changes evaluated from the active (pulse-echo with S=R) and the passive (autocorrelation with S$\neq$R) setup.} 
                    \label{ActVsPass}
        \end{center}
\end{figure}

The noise signals from each source are emitted at consecutive times. To emulate a signal coming from multiple sources at once, the signals recorded from each source $i$ are stacked:

\begin{equation}
h_k(t)= \sum_i G_k(S_i,R,t) \otimes n_i(t),
\end{equation}
where the subscript $k$ holds for temperature.   The velocity variations are then computed using the autocorrelation of this total signal. 
\begin{equation}
C_k(t)= h_k(t)\times h_k(t).
\end{equation}

The velocity variations for each execution $k$ of these experiments are displayed in figure \ref{ActVsPass}, alongside those found with the active experiment. The acquisitions run over about 25 minutes, over which the temperature has increased by about $0.8^{\circ}$C. Within the errors, the velocity variations found with the autocorrelation are the same as those found with the reference GF. \\       

Until now, the analysis was based on the assumption that the autocorrelation used in the passive experiment closely resembles the GF of the medium if there are enough sources, and these sources are stable. In our experiment, the GF is     not reconstructed in the auto-correlation. Nevertheless, figure~\ref{ActVsPass} demonstrates\footnote{Note that the systematic deviation of the passive experiment from the active one  is in part due to a small spatial variability of the temperature change. The auto-correlation (S$\neq$R) is sensitive to the whole medium whereas the pulse-echo wavefield (S=R) is more sensitive to the vicinity of the receiver.} that it is still possible to retrieve correct information about small changes in the medium properties with the resulting auto-correlation. This is a very promising observation that supports the idea that correlation of seismic noise will give the opportunity to monitor weak changes in the Earth with good reliability even when the correlations have not converged to the GF. Indeed, in both \textit{active} and \textit{passive} experiments, we measure the acoustic/seismic signatures of the medium, which naturally include its weak variations.\\

Nevertheless, in order to achieve a proper comparison between our laboratory-scale experiment and seismology, we have to take into account another phenomenon. Indeed, on the earth, the seismic noise sources location is smoothly changing from one week to another. The question addressed in the next section is then: what will happen to our monitoring technique when the background noise is no longer stable, ie. the source distribution changes spatially?

\section{Influence of noise source stability}

\begin{figure}
	\centering
		\includegraphics[width=8cm]{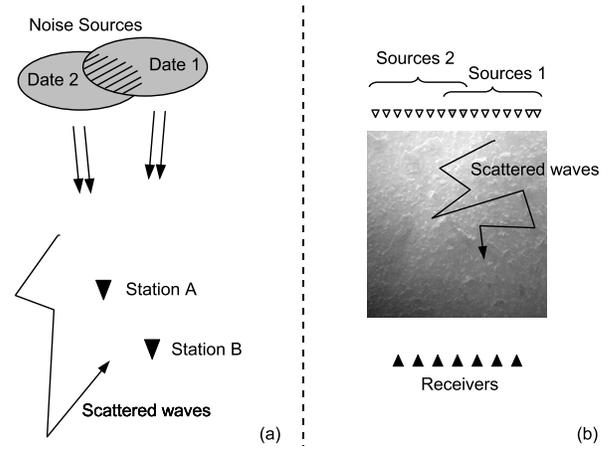}
	\caption{(a) The seismic noise source structure changes from one date to another. (b) Analogous laboratory experiment: two sets of sources are chosen at two different dates.}
	\label{analogy_sources}
\end{figure}

The change in background noise structure is simulated by averaging the autocorrelations for a number of simultaneous uncorrelated sources $i$. The same is done for a slightly different selection of sources a few instants later, then we calculate the relative velocity change $dV/V$ between the two autocorrelations. A simple picture of this is given in Fig.~\ref{analogy_sources}. Imagine for instance that at date $k=$0, sources 1 and 2 are active, and at date $k=$1 sources 2 and 3. For three sources, the decorrelation of the signals can be analyzed theoretically. For the first experiment ($k=$0), the record is:

\begin{equation}
h_0= G(S_1,R,t) \otimes n_1(t) + G(S_2,Rt) \otimes n_2(t)
\end{equation}
and its auto-correlation reads:
\begin{equation}
C_0 = h_0 \times h_0
\end{equation}
or:

\begin{eqnarray}
C_0 & = & G(S_1,R,t) \times G(S_1,R,t) \otimes n_1(t) \times n_1(-t)     \\
&& {} + G(S_2,R,t) \times G(S_2,R,t) \otimes n_2(t) \times n_2(-t) \\
&& {} + 2G(S_1,R,t) \times G(S_2,R,t) \otimes n_1(t) \times n_2(-t)
\end{eqnarray}

Since $n_1(t) \times n_2(t)$ is almost zero, we neglect the third term. 
For simplicity, we shorten the notation:

\begin{equation}
G(S_i,R,t) \times G(S_i,R,t) \otimes n_i(t) \times n_i(-t) = AC_i(t)
\end{equation}
Which leads to :
\begin{equation}
C_0(t)=AC_1(t) + AC_2(t)
\end{equation}

\begin{figure}
	\centering
		\includegraphics[width=8cm]{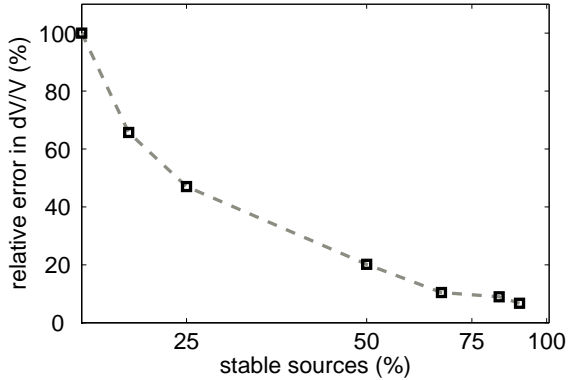}
	\caption{Relative error in the estimation of $\varepsilon=dV/V$, versus the ratio of unchanged-to-total amount of noise sources (x-axis in log. scale).}
	\label{fig12}
\end{figure}

Similarly, for the second experiment ($k=$1), in which the signal is stretched due to the velocity change and the sources used are numbered 2 and 3, we get:

\begin{equation}
C_k = AC_2(t[1- \varepsilon]) + AC_3(t[1- \varepsilon]).
\end{equation}

The objective is to find the $ \varepsilon_k$ that maximizes the cross correlation coefficient $CC_k$ defined in Eq.~\ref{CCk}. The terms $AC_1$ and $AC_3$ are decorrelated waveforms that play the role of fluctuations (eq.~\ref{hk}) and will contribute to the term $B$ in eq.~\ref{eqCCnu}. Assuming that the variance of $AC_k$ is constant: $\langle AC_1^2\rangle =\langle AC_2^2\rangle=\langle AC_3^2\rangle$, we find $A=\frac{1}{2}$ and the standard deviation of $B$ simplifies as $\sqrt{\langle B^2\rangle } = \sqrt{3}\sqrt{2\pi}/{2\sqrt{T\Delta \omega}}$.
From this latter equation, we deduce that a proper estimation of $dV/V$ is carried out if we process a sufficiently large amount of data: $\sqrt{T\Delta \omega}\gg \sqrt{3}\sqrt{2\pi}$. This figure is valid for two sets of sources that have 50\% of sources in common. The same calculation can be carried out for any ratio of unchanged-to-total amount of uncorrelated sources. 

As an example, we report in Fig.~\ref{fig12} the relative experimental error in the estimation of $dV/V$ for various unchanged-to-total source amount ratio. The velocity evolution retrieved through the autocorrelation, as shown in fig \ref{ActVsPass}, is used as a reference. For different ratios of spatially unchanged sources, the velocity changes are computed again, and the deviation with respect to the reference is considered the error. In fig \ref{fig12} the relative error estimation is shown for different ratios of unchanged-to-total sources. When half the sources remain stationary, the relative error is $\sim$ 20 \%, meaning that we have access to a rough (but relevant and interpretable) estimate of the velocity change.
With the given experimental coda duration and bandwidth, we observe that a satisfying estimation of $dV/V$   is obtained if 50\% of the sources are unchanged. \\

In this framework, we conclude more generally that the spatial instability of the source distribution is not a limitation for noise-based correlation monitoring as long as at least part of the noise spatial distribution is stable. The smaller the stable area, the harder the $dV/V$ estimation, and  similar to the conclusion of section III, a larger the integration time $T$ or bandwidth $\Delta \omega$ is required.

\section{Discussion and Perspectives}

In this paper we conducted laboratory experiments with ultrasonic waves to monitor weak velocity changes in the medium. To that end, we employed and compared several procedures that process small phase shift in the diffuse coda waves. These phase shifts correspond to change of arrival time in the waveforms, which were due to small temperature changes in the medium. \\

The paper began with an active (pulse-echo) experiment (section III), in which we compared the \textit{doublet} (or CSMWT) technique to the more recent \textit{stretching} technique. The former is based on Fourier analysis in multiple time windows. The latter is based on the interpolation of the whole waveform and on a grid search optimization. The latter was found to require more computing power, but was also found to be more stable toward noise in the data. In the second part of the paper (section IV), we processed the auto-correlation of noise records acquired in the same medium. Active (pulse-echo) and passive (auto-correlations) data were processed using exactly the same processing procedure: the relative velocity changes of the medium were deduced from the late arrivals using the \textit{stretching} technique. Very similar results were found in both cases, although auto-correlations had not at all converged to the GF. We therefore demonstrated that, contrary to prior belief, passive monitoring with ambient noise remains possible even when the correlation has not converged to the GF. In other words, noise-based monitoring requires weaker assumptions than noise-based imaging. In the last part of the present document (section V), we tested the robustness of the noise-based monitoring technique in the case of unstable distributions of noise sources. We demonstrated that, as long as a certain portion of the sources is stable, velocity variations can still be retrieved. 

Even though   we consider a laboratory experiment in this paper, in practice the results can be extended to different scales, among which seismology is of particular interest. Note that the idea that the coda of the correlations contains precious informations has been recently demonstrated by studying the Correlation of the Coda of the Correlation (C3)\cite{stehly2008}. 
In previous monitoring studies along the San-Andreas faultline at Parkfield, California \cite{brenguier2008b}, there were some doubts as to whether or not the GF was properly reconstructed in the correlations. This was due to i) the imperfect source distribution in the ocean, ii) to the short time-series over which correlations were performed and iii) the low quality data for frequencies below 1~Hz. Nevertheless, it was still possible to observe small variations in the coda of the correlations. From our laboratory experiment, we can confirm that these latter changes are actual physical observations that can be interpreted by velocity changes in the crust. On a somewhat smaller scale, passive monitoring can be applied to seismic prospecting on reservoirs. On such reservoirs, the technique could for instance be used to follow the effect of fluid flows as oil or gas. During production, we indeed expect relative changes of velocity in the reservoir of the order of a few percent\cite{ellison2004}, which seems to be observable by the method presented here. It might also be possible to detect the velocity variations caused by stress changes following subsidence. This monitoring could either be performed with reproducible active sources or with background seismic noise.\\

\section*{Acknowledgments}

This work was partially funded by the University J. Fourier TUNES department, by the CNRS-INSU \textit{AIPI} program, and by the ANR JC08\_313906 \textit{SISDIF} grant. The authors acknowledge R. L. Weaver, A. Verdel and F. Brenguier for scientific discussions and comments, and B. Vial for technical help in designing and setting up the experiment.


\begin{thebibliography}{10}
\newcommand{\enquote}[1]{``#1''}
\expandafter\ifx\csname url\endcsname\relax
  \def\url#1{\texttt{#1}}\fi
\expandafter\ifx\csname urlprefix\endcsname\relax\def\urlprefix{URL }\fi
\providecommand{\bibinfo}[2]{#2}
\providecommand{\noopsort}[1]{}
\providecommand{\switchargs}[2]{#2#1}

\bibitem{duvall1993}
\bibinfo{author}{T.~L. Duvall}, \bibinfo{author}{S.~M. Jefferies},
  \bibinfo{author}{J.~W. Harvey}, and \bibinfo{author}{M.~A. Pomerantz},
  \enquote{\bibinfo{title}{Time-distance helioseismology}},
  \bibinfo{journal}{Nature} \textbf{\bibinfo{volume}{362}},
  \bibinfo{pages}{430--432} (\bibinfo{year}{1993}).

\bibitem{weaver2001}
\bibinfo{author}{R.~L. Weaver} and \bibinfo{author}{O.~I. Lobkis},
  \enquote{\bibinfo{title}{Ultrasonics without a source: Thermal fluctuation
  correlations at {MHz} frequencies}}, \bibinfo{journal}{Phys. Rev. Lett.}
  \textbf{\bibinfo{volume}{87}}, \bibinfo{pages}{134301}
  (\bibinfo{year}{2001}).

\bibitem{lobkis2001}
\bibinfo{author}{O.~I. Lobkis} and \bibinfo{author}{R.~L. Weaver},
  \enquote{\bibinfo{title}{On the emergence of the {Green's} function in the
  correlations of a diffuse field}}, \bibinfo{journal}{J. Acoust. Soc. Am.}
  \textbf{\bibinfo{volume}{110}}, \bibinfo{pages}{3011--3017}
  (\bibinfo{year}{2001}).

\bibitem{larose2006c}
\bibinfo{author}{E.~Larose}, \bibinfo{author}{L.~Margerin},
  \bibinfo{author}{A.~Derode}, \bibinfo{author}{B.~van Tiggelen},
  \bibinfo{author}{M.~Campillo}, \bibinfo{author}{N.~Shapiro},
  \bibinfo{author}{A.~Paul}, \bibinfo{author}{L.~Stehly}, and
  \bibinfo{author}{M.~Tanter}, \enquote{\bibinfo{title}{{Correlation of random
  wavefields: An interdisciplinary review}}}, \bibinfo{journal}{{Geophysics}}
  \textbf{\bibinfo{volume}{{71}}}, \bibinfo{pages}{{SI11--SI21}}
  (\bibinfo{year}{{2006}}).

\bibitem{campillo2003}
\bibinfo{author}{M.~Campillo} and \bibinfo{author}{A.~Paul},
  \enquote{\bibinfo{title}{Long range correlations in the diffuse seismic
  coda}}, \bibinfo{journal}{Science} \textbf{\bibinfo{volume}{299}},
  \bibinfo{pages}{547--549} (\bibinfo{year}{2003}).

\bibitem{shapiro2004}
\bibinfo{author}{N.~M. Shapiro} and \bibinfo{author}{M.~Campillo},
  \enquote{\bibinfo{title}{Emergence of broadband rayleigh waves from
  correlations of the ambient seismic noise}}, \bibinfo{journal}{Geophys. Res.
  Lett.} \textbf{\bibinfo{volume}{31}}, \bibinfo{pages}{L7~614}
  (\bibinfo{year}{2004}).

\bibitem{poupinet1984}
\bibinfo{author}{G.~Poupinet}, \bibinfo{author}{W.~L. Ellsworth}, and
  \bibinfo{author}{J.~Frechet}, \enquote{\bibinfo{title}{Monitoring velocity
  variations in the crust using earthquake doublets: an application to the
  calaveras fault, california}}, \bibinfo{journal}{J. Geophys. Res.}
  \textbf{\bibinfo{volume}{89}}, \bibinfo{pages}{5719--5731}
  (\bibinfo{year}{1984}).

\bibitem{pine1988}
\bibinfo{author}{D.~J. Pine}, \bibinfo{author}{D.~A. Weitz},
  \bibinfo{author}{P.~M. Chaikin}, and \bibinfo{author}{E.~Herbolzheimer},
  \enquote{\bibinfo{title}{Diffusing-wave spectroscopy}},
  \bibinfo{journal}{Phys. Rev. Lett.} \textbf{\bibinfo{volume}{60}},
  \bibinfo{pages}{1134--1137} (\bibinfo{year}{1988}).

\bibitem{snieder2002}
\bibinfo{author}{R.~Snieder}, \bibinfo{author}{A.~Gr\^et},
  \bibinfo{author}{H.~Douma}, and \bibinfo{author}{J.~Scales},
  \enquote{\bibinfo{title}{Coda wave interferometry for estimating nonlinear
  behavior in seismic velocity}}, \bibinfo{journal}{Science}
  \textbf{\bibinfo{volume}{295}}, \bibinfo{pages}{2253--2255}
  (\bibinfo{year}{2002}).

\bibitem{sabra2006}
\bibinfo{author}{K.~G. Sabra}, \bibinfo{author}{P.~Roux},
  \bibinfo{author}{P.~Gerstoft}, \bibinfo{author}{W.~A. Kuperman}, and
  \bibinfo{author}{M.~C. Fehler}, \enquote{\bibinfo{title}{Extracting coherent
  coda arrivals from cross-correlations of long period seismic waves during the
  mount st. helens 2004 eruption}}, \bibinfo{journal}{Geophys. Res. Lett.}
  \textbf{\bibinfo{volume}{33}}, \bibinfo{pages}{L06313}
  (\bibinfo{year}{2006}).

\bibitem{sens-schoenfelder2006}
\bibinfo{author}{C.~Sens-{Sch\"onfelder}} and \bibinfo{author}{U.~C. Wegler},
  \enquote{\bibinfo{title}{Passive image interferometry and seasonal variations
  of seismic velocities at merapi volcano, indonesia.}},
  \bibinfo{journal}{Geophys. Res. Lett.} \textbf{\bibinfo{volume}{33}},
  \bibinfo{pages}{L21302} (\bibinfo{year}{2006}).

\bibitem{sens-schoenfelder2008}
\bibinfo{author}{C.~Sens-Schoenfelder} and \bibinfo{author}{E.~Larose},
  \enquote{\bibinfo{title}{{Temporal changes in the lunar soil from correlation
  of diffuse vibrations}}}, \bibinfo{journal}{Phys. Rev. E}
  \textbf{\bibinfo{volume}{{78}}}, \bibinfo{pages}{{045601}}
  (\bibinfo{year}{{2008}}).

\bibitem{brenguier2008a}
\bibinfo{author}{F.~Brenguier}, \bibinfo{author}{N.~M. Shapiro},
  \bibinfo{author}{M.~Campillo}, \bibinfo{author}{V.~Ferrazzini},
  \bibinfo{author}{Z.~Duputel}, \bibinfo{author}{O.~Coutant}, and
  \bibinfo{author}{A.~Nercessian}, \enquote{\bibinfo{title}{Toward forecasting
  volcanic eruptions using seismic noise}}, \bibinfo{journal}{Nature
  geosciences} \textbf{\bibinfo{volume}{1}}, \bibinfo{pages}{126--130}
  (\bibinfo{year}{2008}).

\bibitem{rhie2004}
\bibinfo{author}{J.~Rhie} and \bibinfo{author}{B.~Romanowicz},
  \enquote{\bibinfo{title}{Excitation of earth's continuous free oscillations
  by atmosphere-ocean-seafloor coupling}}, \bibinfo{journal}{Nature}
  \textbf{\bibinfo{volume}{431}}, \bibinfo{pages}{552--556}
  (\bibinfo{year}{2004}).

\bibitem{stehly2006}
\bibinfo{author}{L.~Stehly}, \bibinfo{author}{N.~Shapiro}, and
  \bibinfo{author}{M.~Campillo}, \enquote{\bibinfo{title}{A study of the
  seismic noise from its long-range correlation properties}},
  \bibinfo{journal}{J. Geophys. Res.} \textbf{\bibinfo{volume}{111}},
  \bibinfo{pages}{B10306} (\bibinfo{year}{2006}).

\bibitem{lobkis2003}
\bibinfo{author}{O.~I. Lobkis} and \bibinfo{author}{R.~L. Weaver},
  \enquote{\bibinfo{title}{Coda-wave interferometry in finite solids: recovery
  of p-to-s conversion rates in an elastodynamic billiard}},
  \bibinfo{journal}{Phys. Rev. Lett.} \textbf{\bibinfo{volume}{90}},
  \bibinfo{pages}{254302} (\bibinfo{year}{2003}).

\bibitem{larose2009}
\bibinfo{author}{E.~Larose} and \bibinfo{author}{S.~Hall},
  \enquote{\bibinfo{title}{Monitoring stress related velocity variation in
  concrete with a $2.10^{-5}$ relative resolution using diffuse ultrasound.}},
  \bibinfo{journal}{J. Acoust. Soc. Am.} \textbf{\bibinfo{volume}{125}},
  \bibinfo{pages}{1853--1856} (\bibinfo{year}{2009}).

\bibitem{fukushima2003}
\bibinfo{author}{Y.~Fukushima}, \bibinfo{author}{O.~Nishizawa},
  \bibinfo{author}{H.~Sato}, and \bibinfo{author}{M.~Ohtake},
  \enquote{\bibinfo{title}{Laboratory study on scattering characteristics of
  shear waves in rock samples}}, \bibinfo{journal}{Bull. Seismol. Soc. Am.}
  \textbf{\bibinfo{volume}{93}}, \bibinfo{pages}{253--263}
  (\bibinfo{year}{2003}).

\bibitem{vanwijk2004}
\bibinfo{author}{K.~{van Wijk}}, \bibinfo{author}{M.~Haney}, and
  \bibinfo{author}{J.~A. Scales}, \enquote{\bibinfo{title}{1d energy transport
  in a strongly scattering laboratory model}}, \bibinfo{journal}{Phys. Rev. E}
  \textbf{\bibinfo{volume}{69}} (\bibinfo{year}{2004}).

\bibitem{larose2005b}
\bibinfo{author}{E.~Larose}, \bibinfo{author}{A.~Derode},
  \bibinfo{author}{D.~Clorennec}, \bibinfo{author}{L.~Margerin}, and
  \bibinfo{author}{M.~Campillo}, \enquote{\bibinfo{title}{Passive retrieval of
  rayleigh waves in disordered elastic media}}, \bibinfo{journal}{Phys. Rev. E}
  \textbf{\bibinfo{volume}{72}}, \bibinfo{pages}{{046607}}
  (\bibinfo{year}{2005}).

\bibitem{zagzebski1996}
\bibinfo{author}{J.~A. Zagzebski}, \emph{\bibinfo{title}{Essentials of
  Ultrasound Physics}} (\bibinfo{publisher}{Mosby Inc}) (\bibinfo{year}{1996}).

\bibitem{brenguier2008b}
\bibinfo{author}{F.~Brenguier}, \bibinfo{author}{M.~Campillo},
  \bibinfo{author}{C.~Hadziioannou}, \bibinfo{author}{N.~M. Shapiro},
  \bibinfo{author}{R.~M. Nadeau}, and \bibinfo{author}{E.~Larose},
  \enquote{\bibinfo{title}{{Postseismic relaxation along the San Andreas fault
  at Parkfield from continuous seismological observations}}},
  \bibinfo{journal}{Science} \textbf{\bibinfo{volume}{{321}}},
  \bibinfo{pages}{{1478--1481}} (\bibinfo{year}{{2008}}).

\bibitem{frechet1989}
\bibinfo{author}{J.~Frechet}, \bibinfo{author}{L.~Martel},
  \bibinfo{author}{L.~Nikolla}, and \bibinfo{author}{G.~Poupinet},
  \enquote{\bibinfo{title}{{Application of the cross-spectral moving-window
  technique (CSMWT) to the seismic monitoring of forced fluid migration in a
  rock mass}}}, \bibinfo{journal}{Int. J. Rock Mech. Min. Sci. \& Geomech.
  Abstr.} \textbf{\bibinfo{volume}{{26}}}, \bibinfo{pages}{{221--233}}
  (\bibinfo{year}{{1989}}).

\bibitem{jenkins1969}
\bibinfo{author}{G.~M. Jenkins} and \bibinfo{author}{D.~G. Watts},
  \emph{\bibinfo{title}{Specral analysis and its applications}}
  (\bibinfo{publisher}{Holden-Day, San Francisco}) (\bibinfo{year}{1969}).

\bibitem{endnote1}
\bibinfo{note}{The stationary assumption is done for the sake of simplicity of
  the calculation. Nevertheless, the main conclusions of the article also apply
  to non-stationary waveforms as decaying coda.}

\bibitem{endnote2}
\bibinfo{note}{See EPAPS document \# for a detailed calculation. For more
  inforamtion on EPAPS, see \protect \textit
  {http://www.aip.org/pubservs/epaps.html}}.


\bibitem{stehly2007b}
\bibinfo{author}{L.~Stehly}, \bibinfo{author}{M.~Campillo}, and
  \bibinfo{author}{N.~M. Shapiro}, \enquote{\bibinfo{title}{{Traveltime
  measurements from noise correlation: stability and detection of instrumental
  time-shifts}}}, \bibinfo{journal}{Geophys. J. Int.}
  \textbf{\bibinfo{volume}{{171}}}, \bibinfo{pages}{{223--230}}
  (\bibinfo{year}{{2007}}).

\bibitem{endnote3}
\bibinfo{note}{Note that the systematic deviation of the passive experiment
  from the active one is in part due to a small spatial variability of the
  temperature change. The auto-correlation (S$\not =$R) is sensitive to the
  whole medium whereas the pulse-echo wavefield (S=R) is more sensitive to the
  vicinity of the receiver.}

\bibitem{stehly2008}
\bibinfo{author}{L.~Stehly}, \bibinfo{author}{M.~Campillo},
  \bibinfo{author}{B.~Froment}, and \bibinfo{author}{R.~L. Weaver},
  \enquote{\bibinfo{title}{Reconstructing green's function by correlation of
  the coda of the correlation (c$^3$) of ambient seismic noise}},
  \bibinfo{journal}{J. Geophys. Res.} \textbf{\bibinfo{volume}{113}},
  \bibinfo{pages}{B11306} (\bibinfo{year}{2008}).

\bibitem{ellison2004}
\bibinfo{author}{S.~J. Ellison}, \bibinfo{author}{M.~G. Imhofz},
  \bibinfo{author}{C.~{\c C}oruhz}, \bibinfo{author}{A.~D. Fuqua}, and
  \bibinfo{author}{S.~C. Henry}, \enquote{\bibinfo{title}{Modeling
  offset-dependent reflectivity for time-lapse monitoring of water-flood
  production in thin-layered reservoirs}}, \bibinfo{journal}{Geophysics}
  \textbf{\bibinfo{volume}{69}}, \bibinfo{pages}{25--36}
  (\bibinfo{year}{2004}).

\end{thebibliography}

\end{document}